\documentclass[a4paper,nofootinbib]{revtex4}

\usepackage{amsmath,amssymb,xspace,hyperref,comment}
\usepackage{graphicx}
\usepackage{color}

\newcommand{\lvdW}{\ensuremath{\ell_{\text{vdW}}}\xspace}

\begin{document}
\title{Quantum unbinding 
near a zero temperature liquid-gas transition}
\author{Wilhelm Zwerger}
\affiliation{Technische Universit\"at M\"unchen, Physik Department, James-Franck-Strasse, 85748 Garching, Germany}

\begin{abstract}
We discuss the quantum phase transition from a liquid to a gaseous ground 
state in a Bose fluid with increasing strength of the zero point motion. It is shown
that in the zero pressure limit, the two different ground states are separated by 
a quantum tricritical point whose position is determined by a vanishing two-body 
scattering length. In the presence of a finite three-body scattering amplitude,
the superfluid gas at this point exhibits sound modes whose velocity 
scales linearly with density while the compressibility diverges $\sim p^{-1/3}$ 
in the limit of vanishing pressure $p$. In the liquid regime of 
negative scattering lengths, it is shown that $N$-body bound states exist up to 
arbitrary $N$, consistent with a theorem by Seiringer. The asymptotic scaling 
$a_{-}(N)\sim N^{-1/2}$ of the scattering lengths where they appear from
the continuum is determined from a finite size scaling analysis in
the vicinity of the quantum tricritical point. This also provides a qualitative 
understanding of numerical results for the quantum unbinding of small clusters. 
\end{abstract}
\maketitle

\section{Introduction}

\begin{figure}[t]	
\includegraphics[width=65mm]{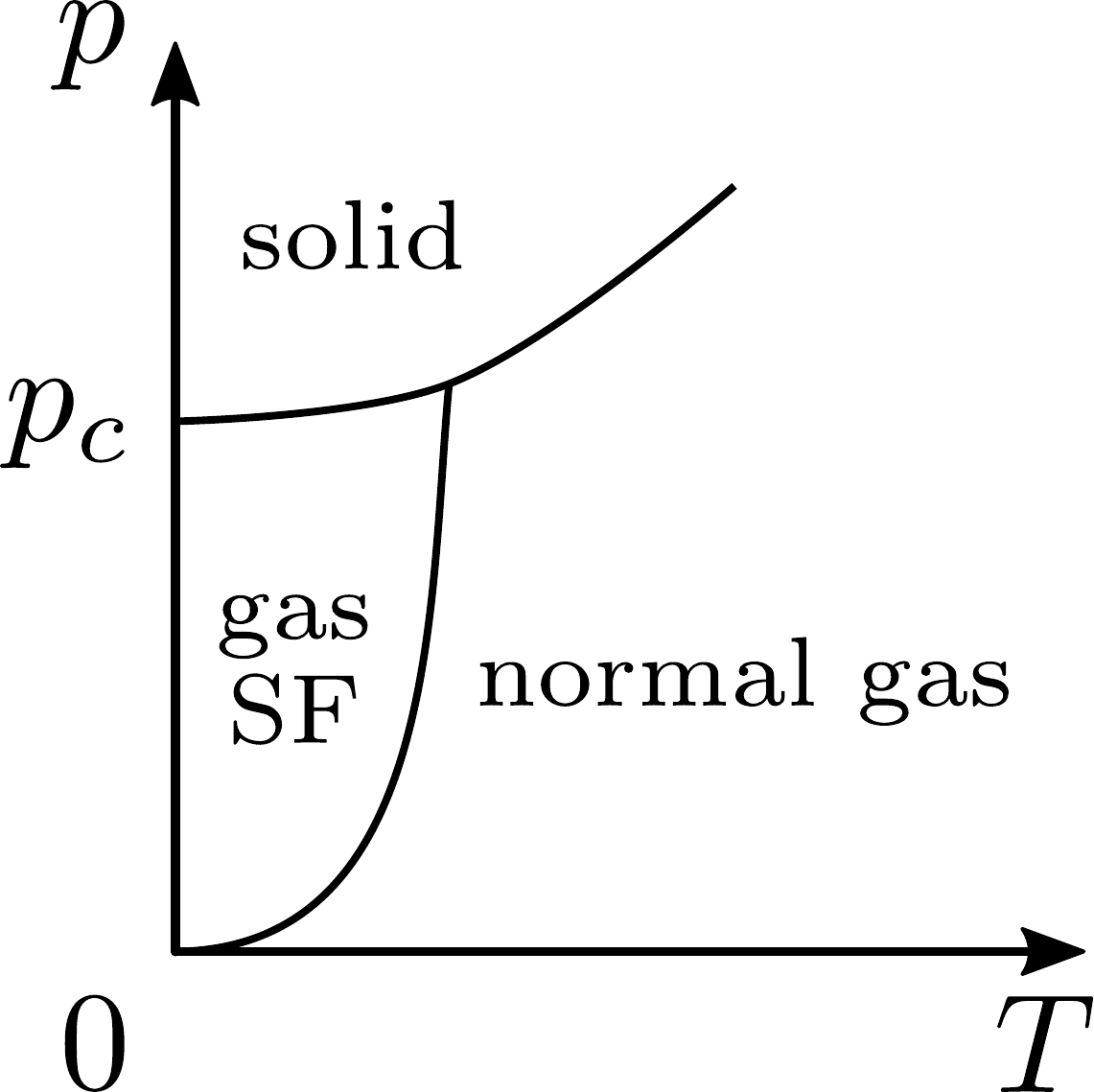}
\caption{A qualitative pressure-temperature phase diagram for a Bose system in the 
regime $\Lambda_{\rm dB}>\Lambda_{\rm dB}^c$, where the ground state below a critical 
pressure $p_c$ is a superfluid gas. Due to $n(p)=\sqrt{2p/g}$ at zero temperature and $p\to 0$,
the transition from the superfluid to the normal gas asymptotically exhibits
a cubic dependence $p(T)\simeq g/\lambda_T^6\sim g\, T^3$.}
\label{fig:phase-diagram}
\end{figure}

It is an empirical fact that the generic equilibrium state of matter at low temperatures 
is a solid. The only exceptions are the two isotopes of Helium
which remain in a liquid phase down to zero temperature. On a microscopic level, 
this is a consequence of zero point motion, which plays an only minor role in most solids.
In Helium, however, it is large enough to destroy crystalline order of
the ground state below a critical external pressure $p_c$.
For particles obeying Bose statistics, which will be considered
here, the phase diagram can be determined in quantitative terms within a fully 
microscopic approach even for a strongly interacting system like 
$^4$He~\cite{cepe95RMP}. The approach also reveals the importance
of Bose statistics in the stability of a liquid state at low temperatures 
and pressure. Indeed, as shown by~\textcite{boni12}, distinguishable particles with the same
amount of zero point energy would actually form a crystal even at $p<p_c$ unless the 
temperature is close to zero. 
The aim of the present work is to investigate a scenario in which the 
zero point energy is even larger than the one in $^4$He. The ground state 
is then eventually a superfluid gas and the resulting finite temperature phase diagram, 
shown schematically in Fig.~\ref{fig:phase-diagram},  exhibits neither a triple- nor a critical point.  
In practice, such a scenario is realized only in the case of spin-polarized 
hydrogen. As a long lived metastable configuration, however, a gaseous state is
also present in dilute, ultracold Bose gases.
Some of the results found here are thus expected to be of
relevance also in this context. Note that for Bose systems which are liquid or
gaseous with a uniform density, the ground state is a superfluid by a rather general 
argument due to~\textcite{legg73helium}.\\

In order to understand the quantum phase transition between a liquid and a gaseous 
ground state with increasing strength of the zero point motion, 
we follow an approach due  to~\textcite{sach11book} and consider the transition out of the
vacuum state into one with a finite particle density $n$ as a function of
the chemical potential $\mu$. In the case where the ground state is a gas,
the associated effective field theory is the well known $\psi^4$-theory for
a complex scalar field. Specifically, the finite density gas with $n(\mu)=\mu/g+\ldots$
is separated from the vacuum at $\mu<0$ by a line of fixed points with infinite 
correlation length. For a liquid ground state, in turn, the
transition out of the vacuum appears at a negative chemical potential $\mu_c<0$. 
It is of first order, i.e. the density jumps from zero to a finite value $\bar{n}$ at
$\mu=\mu_c^+$. The associated line of critical points thus has a finite 
correlation length which approaches infinity, however, close to the point where $\mu_c$ reaches zero.
Microscopically, the location of the quantum tricritical point which separates the liquid and gaseous  
ground states is determined by the condition  $g\!=\!4\pi\hbar^2a/m\equiv 0$ of a vanishing two-body scattering length.
At this point and also in the liquid regime at small negative scattering lengths, the
Bose system is stabilized by three-body interactions. The effective field
theory thus features a $\psi^6$-term as in the standard description of tricritical points.
Expressed in terms of the de Boer parameter  $\Lambda_{\rm dB}$, which measures the
strength of the zero point motion~\cite{debo48},  the ground state at vanishing pressure changes
from a liquid to a gas at a critical value $\Lambda_{\rm dB}^c\simeq 0.7$. For a finite
particle number $N$, the quantum unbinding transition
is shifted to lower values $\Lambda_{\rm dB}^{\ast}(N)<\Lambda_{\rm dB}^c$ 
of the de Boer parameter. This shift can be understood within a finite size scaling analysis in 
the vicinity of the quantum tricritical point, where both the bulk and the surface energy vanish, providing
a qualitative understanding of numerical results for the unbinding of small clusters~\cite{meie96,sevr10}.\\

In standard ultracold Bose gases, the de Boer parameter
is much less than one and thus the interaction supports a large number
of  two-body bound states. At very low densities, however,
they are of no relevance since they are inaccessible kinematically with just two-body 
collisions. In particular, Bose gases with a positive scattering length form
an essentially stable system down to zero temperature. For negative scattering
lengths, in turn, stability of the ground state may be preserved only for small enough 
particle numbers $N<N_c\simeq 0.6\,\ell_0/|a|$ and in the presence of a harmonic trap
with an associated oscillator length $\ell_0$. At finite temperatures, however, an essentially homogeneous
Bose gas with attractive interactions can be studied in a moderately degenerate, non-condensed 
regime $n\lambda_T^3\lesssim 1$, even if the magnitude  $|a|$ of the scattering length is 
increased to values of the order of the average interparticle spacing $n^{-1/3}$ or the thermal 
wavelength $\lambda_T$ by the use of Feshbach resonances~\cite{chin10feshbach}.
This allows to explore basic few-body phenomena like the existence of three-body bound states for
identical bosons, which were predicted by~\textcite{efim70} in a nuclear physics context.
Efimov states were first observed in an ultracold gas of Cesium atoms,
where the formation of a trimer bound state near zero energy shows up as a pronounced 
maximum in the three-body loss coefficient~\cite{krae06efimov}.
Theoretically, one expects an infinite sequence of trimers which may be thought
of as excited states of the lowest lying $3$-body bound state. They
emerge from the two-particle continuum at a sequence of increasingly more
negative scattering lengths $a_{-}^{(n)}(3)$ which approach $a=\pm\infty$ in a 
geometric manner. Near this accumulation point, the effective range of the interactions 
is irrelevant. 
For $n\gg 1$, therefore, the series displays universal behavior 
$a_{-}^{(n+1)}(3)/a_{-}^{(n)}(3)\to 22.69\ldots$ which is associated with
an underlying renormalization group limit cycle~\cite{braa06}.
In practice, only the lowest or possibly the second lowest~\cite{huan14efimov}
of these trimers are accessible,
where universality is violated due to the finite range of the interactions. As shown by~\textcite{schm12efimov},
this leads e.g. to a ratio $a_{-}^{(1)}(3)/a_{-}^{(0)}(3)\simeq 17$, much smaller than in the universal limit.
Many-body bound states exist also for larger particle numbers $N>3$. This has been studied in detail 
for $N=4$, where theory predicts an infinite sequence of
two tetramer states per Efimov trimer~\cite{hamm07,stec09tetramers,schm10tetramers,delt12}. 
Experimentally, the lowest tetramer state has been observed by~\textcite{ferl09} at 
$a_{-}(4)\simeq 0.47\, a_{-}(3)$ and even signatures of a five-body bound state have been inferred
from a characteristic feature in the recombination rate of Cesium near a scattering length $a_{-}(5)\simeq 0.64\,a_{-}(4)$~\cite{zene13}. 
In the present work, we will restrict ourselves to the energetically lowest $N$-body bound states. 
The magnitude of the scattering lengths $a_{-}(N)<0$ where they detach from the continuum form a 
sequence which apparently approaches zero in a monotonic manner. 
 This has been investigated by von Stecher via numerical solutions of the Schr\"odinger equation 
 up to $N=13$~\cite{stec10N-body}. In particular, it turns out 
 that the consecutive ratios $a_{-}(4)/a_{-}(3)\simeq 0.44\; , \; a_{-}(5)/a_{-}(4)\simeq 0.64$
 and  $a_{-}(6)/a_{-}(5)\simeq 0.73$ are not very sensitive to the detailed form of the two-body interactions~\cite{stec11six-body}.\\
   
  A natural question which arises in this context is whether the sequence of $N$-body bound states
 continues up to $N\!=\!\infty$. As will be discussed below, this is indeed the case, at least 
near the quantum tricritical point, where the interaction changes from an overall attractive to a repulsive 
behavior and there is no longer any two-body bound state. The existence of an infinite
sequence of $N$-body bound states with an accumulation point at $a=0$ is
consistent with a theorem due to~\textcite{seir12}, which states that {\bf some} $N$-body bound state 
must exist for arbitrary small negative scattering lengths. Specifically, it turns out that the  
asymptotic dependence $|a_{-}(N)|\sim N^{-1/2}$ of the magnitude of the scattering lengths below which no 
$N$-body bound states exist, is determined by a finite size scaling analysis in the vicinity of the quantum tricritical point. 
It is important to note that these results rely crucially on the assumption of
two-body interactions which have a strong repulsive part at short distances and a finite range,
guaranteeing thermodynamic stability and a stable liquid ground state as in $^4$He. 
By contrast, within zero range or purely attractive interactions, 
the issue of $N$-body bound states is ill-defined for large $N$. 
An example is provided by a Bose gas with zero range attractive interactions 
in two dimensions. As shown by~\textcite{hamm04}, it exhibits an 
infinite sequence of $N$-body bound states. Due to the logarithmic increase of the effective
interaction strength at large distances, their binding energies $B_N$ increase exponentially 
according to $B_{N+1}/B_N\to 8.567\ldots$ while their size $R_N$ goes to zero as $R_{N+1}/R_N=0.3417\ldots$
(see also the recent work by~\textcite{petr18}).
For large $N$, therefore, the behavior of the interaction at short distance becomes relevant.
An exceptional situation appears in one dimension, where the quantum dissociation of a Luttinger liquid into a gaseous 
phase has been discussed by~\textcite{kolo03}. In this case, an attractive Bose gas with zero range interactions 
exhibits a finite temperature liquid-gas transition as shown by~\textcite{herz14}. \\

The paper is structured as follows: In section 2, we introduce the microscopic
model and discuss the resulting two- and three-body problem. 
For vanishing scattering length and in a regime, where no two-body bound state exists,
a repulsive three-body interaction arises from a positive value
of the associated scattering hypervolume introduced by~\textcite{tan08bose}. 
An effective field theory description of the zero temperature liquid-gas transition
in the many-body system near vanishing scattering length is developed in section 3, 
extending the early numerical approach to this problem by~\textcite{mill77}.   
Depending on the sign of $a$, the finite density superfluid 
is separated from the vacuum state by an either first or a second order line of fixed points 
which meet at a quantum tricritical point. Its position 
is determined by the condition of vanishing scattering length, i.e. by two-body physics. 
Moreover, in section 4, we address the problem for finite particle numbers. It is
shown that the threshold scattering lengths $a_{-}(N)$ for the
existence of $N$-body bound states in the limit $N\gg 1$ are determined by a finite size
scaling analysis in the vicinity of the quantum tricritical point, which separates the 
liquid and gaseous ground states of the bulk system.   
A conclusion and a discussion of open problems is presented in section 5.

\section{Microscopic model and few-body physics}

We consider a generic Hamiltonian for a system of bosons with pure two-body interactions.
The associated first quantized Hamiltonian
\begin{equation}
\hat{H}_N=-\frac{\hbar^2}{2m}\sum_{i=1}^N\nabla^2_i \, +\! \sum_{1\leq i<j\leq N} V(\mathbf{x}_i-\mathbf{x}_j)
 \label{eq:Hamiltonian}
 \end{equation}
gives rise to a proper thermodynamics 
provided the interaction fulfills certain conditions. Specifically, as shown by~\textcite{fish64stability},
a sufficient condition for the existence of a well defined thermodynamic limit is that the two-body potential
$V(r)\geq -\epsilon$ has a finite lower bound, decays faster than $1/r^3$ at large distances and increases 
more rapidly than $1/r^3$ for separations smaller than a short range scale $\sigma$.
Interactions of this type guarantee stability of the many-body problem since they obey
\begin{equation}
\sum_{1\leq i<j\leq N} V(\mathbf{x}_i-\mathbf{x}_j) > -B\cdot N
 \label{eq:stability}
 \end{equation}
for all possible configurations with a positive constant $B$, which depends on $\epsilon$ but not on other 
variables like density, which are specific for different ground states.
The characteristic energy and length scales $\epsilon$ and $\sigma$ determine the de Boer parameter
\begin{equation}
\Lambda_{\rm dB}=\frac{\hbar}{\sigma\sqrt{m\epsilon}}
 \label{eq:deBoer}
 \end{equation} 
 as the square root of the ratio between the zero point energy on the scale $\sigma$ and the depth $\epsilon$ of
 the attractive part of the potential.  More specifically, we consider 
interactions with an asymptotic van der Waals tail $V(r\to\infty)=-C_6/r^6$ and a strong short range repulsion
at $r\lesssim\sigma$ whose detailed behavior is not important except for the constraint that the 
potential  increases more rapidly than $1/r^3$. A standard example is provided by the
Lennard-Jones potential $ V(r)=4\epsilon\, [ \left(\sigma/r\right)^{12} - \left(\sigma/r\right)^6 ]$
where the short distance scale $\sigma$ and the depth $\epsilon$ are connected with the strength of
the van der Waals tail via $C_6=4\epsilon\,\sigma^6$. Independent of their
precise form, the class of potentials which obey Eq.~(\ref{eq:stability}) lead to a well defined many-body ground state, whose
energy $E_0(N)=u\, N+\ldots $ scales linearly with the particle number. 
In the regime $\Lambda_{\rm dB}\!>\!\Lambda_{\rm dB}^c$ of a gaseous ground state,
the repulsive part of the interaction dominates and the scattering length $a$ is positive.
Introducing $g=4\pi\hbar^2 a/m>0$, both the energy per particle $u(n)\!\to\! gn/2=\sqrt{gp/2}$
and the density $n(p)\!\to\!\sqrt{2p/g}$ then vanish in the zero pressure limit. By contrast, a liquid ground state
has a finite density $\bar{n}$ at vanishing pressure. As a result, $u(\bar{n})$ is finite and 
negative. Specifically, this is the case for $^4$He, where $\Lambda_{\rm dB}\simeq 0.4$. The attractive part of the two-body 
interaction is then just barely sufficient to give rise to a bound state with a binding energy $B_2\simeq k_B\cdot 1.7\,{\rm mK}$. 
It is tiny compared with the ground state energy per particle $u\simeq - k_B\cdot 7\,{\rm K}$ in the bulk liquid 
at zero pressure and dimensionless density $\bar{n}\sigma^3\simeq 0.364$~\cite{cepe95RMP,kalo81}.\\

\begin{figure}[t]	
\includegraphics[width=65mm]{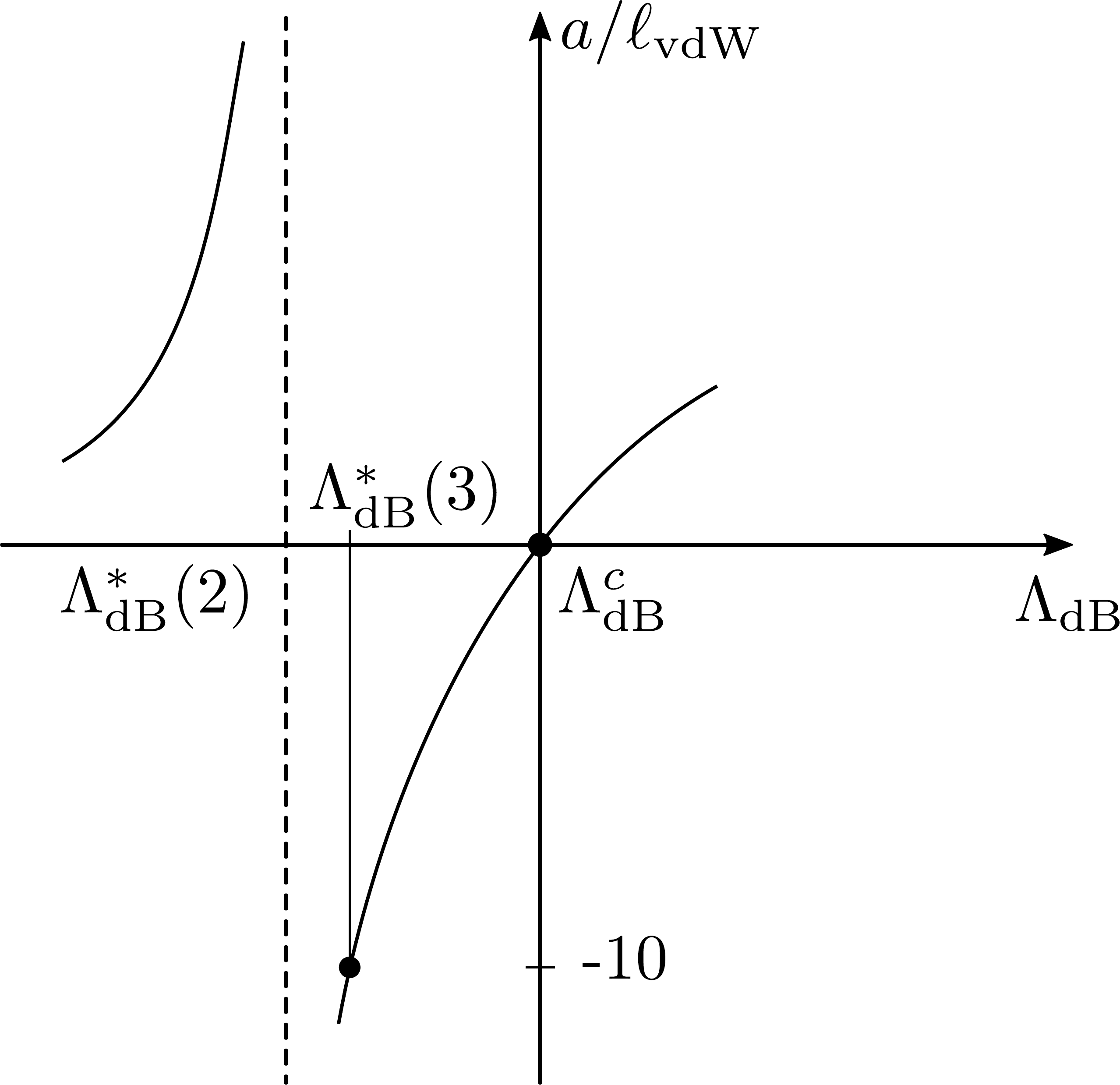}
\caption{Qualitative dependence of the scattering length in units of the van der Waals length $\lvdW$
as a function of the de Boer parameter defined in Eq.~(\ref{eq:deBoer}). The last two-body bound state 
disappears beyond the pole of the scattering length at $\Lambda_{\rm dB}^{\ast}(N=2)$ indicated by the 
dashed vertical line. The scattering length reaches zero at a critical value $\Lambda_{\rm dB}^c\simeq 0.7$,
beyond which it stays positive. The value $\Lambda_{\rm dB}^{\ast}(N=3)$ for the disappearance of
three-body bound states is also indicated.}
\label{fig:deBoer}
\end{figure}

At the level of just two particles, determining the spectrum of $\hat{H}_2$ is an elementary problem in 
quantum mechanics. In the standard regime $\Lambda_{\rm dB}\ll 1$, there is a large 
number $N_b\simeq 1/(\pi\Lambda_{\rm dB})\gg 1$ of s-wave bound states. Upon reduction of the 
strength of the attractive interaction, their number decreases 
and eventually reaches zero at a critical value of the de Boer parameter. In physical terms, this happens 
when the van der Waals length $\lvdW=(mC_6/\hbar^2)^{1/4}/2$,  which is determined solely by the asymptotic form $-C_6/r^6$ 
of the interatomic interaction, has decreased to a value of the order of the short distance scale $\sigma$. 
For the specific case of a Lennard-Jones potential, where the scattering length can be determined with high precision 
(see e.g.~\textcite{gome12LJ-potential}), the limit beyond which the two-body Hamiltonian $\hat{H}_2$ no longer has 
a bound state is reached at $\Lambda_{\rm dB}^{\ast}(N=2)=0.423\ldots$ or $\ell_{\rm vdW}=1.09\, \sigma$. At this 
point, the scattering length jumps form $+\infty$ to $-\infty$ and the last two-body bound state unbinds.  
Upon further increasing the de Boer parameter, the scattering length increases monotonically from
$-\infty$ towards zero, which is reached at some critical value $\Lambda_{\rm dB}^c$. Specifically, one finds 
$\Lambda_{\rm dB}^c=0.679\ldots$ for a Lennard-Jones potential, corresponding to a van der Waals length 
$\ell_{\rm vdW}\vert_c\simeq 0.86\,\sigma$. Increasing $\Lambda_{\rm dB}$ 
beyond its critical value, the interaction turns into a dominantly repulsive one, and the scattering length stays positive. 
An analytical expression is obtained for the pure $4\epsilon\left(\sigma/r\right)^{12}$ part of the Lennard-Jones potential, where
\footnote{In practice, the limit $\Lambda_{\rm dB}\gg 1$ cannot be reached since the maximum attainable de Boer parameter 
is around $\Lambda_{\rm dB}\simeq 0.74$, which is the value estimated for spin-polarized hydrogen, see~\textcite{stwa76nosanov}.}
\begin{equation}
a_{\rm LJ}(\Lambda_{\rm dB}\gg 1)=\lvdW\, \frac{\Gamma(0.9)}{\Gamma(1.1)}\left( 8 \Lambda_{\rm dB}^{3/2}/5\right)^{1/5}
 \label{eq:scattering-length}
 \end{equation}  
which is of order $\lvdW\simeq\sigma$ for realistic values of $\Lambda_{\rm dB}\simeq 1$. 
The qualitative dependence of the scattering length on the de Boer parameter in the regime of interest 
is sketched in Fig.~\ref{fig:deBoer}. In the vicinity of the last zero crossing near $\Lambda_{\rm dB}^c$, 
the scattering length vanishes linearly  
\begin{equation}
a(\Lambda_{\rm dB})=a_{\Lambda}\,\lvdW\left(\Lambda_{\rm dB}-\Lambda_{\rm dB}^c\right)+\ldots
 \label{eq:zero-crossing}
 \end{equation}   
with a numerical constant $a_{\Lambda}$ of order one. This result is expected to hold quite generally for 
the class of potentials considered here. \\

The three-body problem for short range potentials with a van der Waals tail can be solved numerically.
For the specific case of a Lennard-Jones interaction and in the regime of a shallow potential well,
where no two-body bound state is present, this has been investigated by~\textcite{mest17} using the 
hyperspherical approach. It turns out that the last three-body bound state disappears beyond a critical 
value $\Lambda_{\rm dB}^{\ast}(3)\simeq 0.45$ of the Boer parameter.
Expressed in terms of the scattering length, this corresponds to $a_{-}(3)=-9.6\,\lvdW$~\cite{mest17}.  
The order of magnitude $a_{-}(3)/\lvdW\simeq -9$ of this ratio
is generic for interactions involving a van der Waals tail at large distances, independent
of the short range scale $\sigma$.  
For single channel potentials with a large number $N_b\gg 1$ of s-wave bound states,
the ratio $(a_{-}(3)/\lvdW)\vert_{N_b\gg 1}=-9.45$ has in fact a universal value 
in the absence of three-body forces, as shown by~\textcite{wang12}.  
In practice, a change in the scattering length relies on the use of Feshbach resonances.
A nearly universal value of the ratio $a_{-}(3)/\lvdW\simeq -9$ then appears only in the open-channel 
dominated limit~\cite{schm12efimov}. Moreover, even in this regime, the experimentally observed 
ratio $a_{-}(3)/\lvdW$ varies within a range between $-8$ and $-10$, a spread which is likely to be 
caused by different strengths of repulsive short-range three-body forces, see~\textcite{lang18}.  \\

As discussed in the introduction, bound states exist also for larger particle numbers. 
Based on the numerical results obtained by von Stecher~\cite{stec10N-body,stec11six-body},
the magnitudes $|a_{-}(N)|$ of the threshold scattering lengths below which no $N$-body bound 
state exists, form a monotonically decreasing sequence.
By Seiringer's theorem mentioned above, this sequence has $a=0$ as a lower limit which,
as will be shown in section IV below, is approached in a power law form as $N\gg 1$.
For a study of $N$-body bound states in the limit of large $N$, 
it is therefore sufficient to consider the regime of small, negative scattering lengths $|a|\ll \lvdW\simeq\sigma$.
Fine tuning of the interaction to the special value $\Lambda_{\rm dB}=\Lambda_{\rm dB}^c$, where the scattering 
length vanishes, is naively expected to realize an ideal Bose gas.
This is not true, however, because a vanishing scattering length does not
imply that the interactions have disappeared completely. 
Indeed, two-body interactions are still present in relative angular momenta $l=2,4\ldots$.
For interactions with an attractive van der Waals behavior $-C_6/r^6$, 
the leading d-wave contribution to the two-body scattering amplitude $f(k,\theta)$ 
for identical bosons at low relative momenta $k$ is
\footnote{Note that the van der Waals interaction decays too slowly to guarantee the standard power law 
$f_l(k\to 0)\sim k^{2l}$ for the behavior of the l-wave scattering amplitude at low momenta in the leading $l=2$ contribution.} 
\begin{equation}
 \left.  f(k,\theta)\right |_{a=0}=\frac{64\pi}{63}\,\lvdW \left(k\lvdW\right)^3\,P_2(\cos{\theta}) +\ldots
 \label{eq:d-wave}
 \end{equation}  
 In order to understand in more detail the limit of bosons
with nearly vanishing scattering length, it is instructive to consider the two- and three-body 
problems in a cubic box of size $L$ with periodic boundary conditions. For just two particles,
the three leading terms of the associated ground state energy
\begin{equation}
E_0(2\vert L)=\frac{g}{L^3}\left[ 1+ 2.837\ldots\left(\frac{a}{L}\right) + 6.375\ldots \left(\frac{a}{L}\right)^2\right] +\mathcal{O}(L^{-6})
 \label{eq:Luescher}
 \end{equation}  
only depend on the scattering length, as was shown by~\textcite{lues86}
in the context of finite size scaling for quantum field theories on a lattice. If the scattering length
is zero, the first non-vanishing contribution to the two-particle ground state energy therefore scales
like $L^{-6}$, i.e. like the {\bf square} of the effective density $n\simeq 1/L^3$. In fact, such a dependence
appears as the leading term at $a=0$ in the extension of the result~(\ref{eq:Luescher}) to three particles,
which has been derived by~\textcite{tan08bose}. Specifically, the ground state energy of three bosons with 
vanishing two-body scattering length in a box with periodic boundary conditions turns out to be~\cite{tan08bose} 
\begin{equation}
\left. E_0(3\vert L)\right |_{a=0}=\frac{\hbar^2 D}{mL^6} +\mathcal{O}(L^{-8})\, .
 \label{eq:Tan}
 \end{equation} 
 The associated parameter $D$ has been called the three-body scattering hypervolume 
by~\textcite{tan08bose}. It has dimension (length)$^4$ and characterizes the
strength of three-body interactions. In analogy to the definition of the scattering length
via the asymptotic behavior $\psi_{E=0}(\mathbf{x}_1,\mathbf{x}_2)=1-a/r_{12}$ of the
two-body wave function at zero energy, the three-body scattering hypervolume $D$ 
can alternatively be defined by the asymptotic behavior
\begin{equation}
\psi_{E=0}(\mathbf{x}_1,\mathbf{x}_2, \mathbf{x}_3)\vert_{a=0}=1-\frac{\sqrt{3}\, D}{2\pi^3 (r_{12}^2+r_{13}^2+r_{23}^2)^2} \; +\ldots
 \label{eq:hypervolume}
 \end{equation} 
 of the three-body wave function at zero energy and vanishing scattering length~\cite{tan08bose}. 
Moreover, the position of the three-body bound states is determined by the poles of the hypervolume $D$. 
Decreasing the de Boer parameter from its critical value $\Lambda_{\rm dB}^c$ to $\Lambda_{\rm dB}^{\ast}(3)$,      
the three-body scattering hypervolume $D$ thus approaches minus infinity. In the regime 
$\Lambda_{\rm dB}<\Lambda_{\rm dB}^{\ast}(3)$ where three-body bound states exist,
it also acquires a finite imaginary part. Indeed, as shown by~\textcite{zhu17}, the imaginary part of 
$D$ describes the three-body loss rate via 
\begin{equation}
\label{eq:Bose-lifetime}
\Gamma_3=-\dot{N}_{\rm 3-body} /N\, =L_3\, n^2 = -\frac{\hbar}{m}\, {\rm Im} (D)\, n^2\, .
\end{equation}

In the following, we are interested in the regime 
$\Lambda_{\rm dB}^{\ast}(3)< \Lambda_{\rm dB}\lesssim \Lambda_{\rm dB}^c$ 
close to vanishing scattering length, where no three-body bound states exist and 
the parameter $D$ is real. Our fundamental assumption is that $D>0$ is positive 
in a finite range of scattering lengths near $a=0$, where it must be of order $\sigma^4\simeq \lvdW^4$ 
by dimensional analysis. This assumption is supported by the numerical results 
for the many-body problem by~\textcite{mill77}, as will be discussed in more detail 
in the context of Eq.~(\ref{eq:curvature}) below. It is also consistent with an explicit calculation of 
the three-body scattering hypervolume for a Gaussian potential
$V(r)\sim \upsilon_0\exp{(-r^2/r_0^2)}$ by~\textcite{zhu17}
\footnote{see also the related recent work by~\textcite{mest19} for a square well potential.}. 
Indeed, they find $D>0$ not only in the purely repulsive case $\upsilon_0>0$ but also in a finite range 
of attractive interactions as long as the magnitude of the associated 
negative scattering length obeys $|a|\lesssim r_0$. Since their interaction
vanishes identically at $a=0$, one trivially finds $D=0$ at the zero crossing 
of the scattering length. This is an artefact, however, of an interaction which is 
either purely repulsive or attractive and is not expected to apply for potentials e.g. of 
the Lennard Jones type near $\Lambda_{\rm dB}^c$. It is important to note that
a positive value of $D$ near vanishing scattering length may also arise 
from repulsive three-body forces which were studied in connection with the unexpectedly wide range of the observed ratio 
$a_{-}(3)/\lvdW$ by~\textcite{lang18}. The property $D(a=0)>0$ is not guaranteed however, 
for the full class of two-body interactions which obey the stability criterion~(\ref{eq:stability}).   
Indeed, as shown by~\textcite{baum97}, it is possible to construct interactions which obey~(\ref{eq:stability}) 
but they exhibit a three-body bound state even in the regime of positive scattering lengths.
In contrast to the expected gaseous BEC, which is indeed found 
in the physically relevant example of spin-polarized hydrogen~\cite{frie98},
the  non-monotonic interactions considered by Baumgartner lead to a crystalline ground state. 
In practice, this is the case for $\Lambda_{\rm dB}<\Lambda_{\rm dB}^{\rm c, solid}\simeq 0.37$~\cite{nosa75}.
By contrast, here we consider the situation near $\Lambda_{\rm dB}^c\simeq 0.7$ where the ground state 
above  $\Lambda_{\rm dB}^c$ is a gas. A positive scattering length in the absence of any two-body bound state 
then implies the absence of bound states for all $N>2$.\\

A crucial consequence of the assumption of a positive value $D>0$ of the three-body scattering 
hypervolume near $a=0$ is that, at finite density,
the energy per particle $u(n)=(\hbar^2 D/6m)\cdot n^2$ 
scales quadratically with $n$~\cite{tan08bose}.  At vanishing scattering length, therefore,
the many-body Bose fluid is stabilized by repulsive three-body interactions. 
The relation between pressure and chemical 
potential right at $a=0$ is then of the form
\begin{equation}
\left. p(\mu)\right |_{a=0}=\left(\frac{8m}{9\hbar^2 D}\right)^{1/2}\!\cdot\mu^{3/2} \;\;\; \to \;\;\; \left. \mu(n) \right |_{a=0}=\frac{\hbar^2 D}{2m}\cdot n^2\, .
 \label{eq:critical-pressure}
 \end{equation}  
 As a result, the density $n(\mu)=\partial p/\partial\mu$ scales with the square root of the 
 chemical potential rather than the linear behavior found for positive scattering lengths. 
 This is a consequence of the non-standard critical exponent $\beta=1/4$ 
 associated with the appearance of a finite order parameter $|\psi|(\mu)\sim\mu^{\beta}$
 at a {\bf tri}critical point, which will be discussed in more detail below. 
 The associated compressibility $\tilde{\kappa}= \partial n/\partial\mu$ 
exhibits a weak power law divergence  
 \begin{equation}
\left. \tilde{\kappa}(p)\right |_{a=0}=3\left(\frac{m}{9\hbar^2 D}\right)^{2/3}\cdot p^{-1/3}
 \label{eq:critical-kappa}
  \end{equation}
in the limit of vanishing pressure, instead of approaching a constant $\tilde{\kappa}(p\to 0)=1/g$  
for finite, positive values of the scattering length. 
The relation~(\ref{eq:critical-kappa}) provides a possible experimental tool to verify the presence of 
 three-body interactions in a Bose gas at vanishing scattering length and to determine 
 the associated three-body scattering hyper\-volume $D$. Indeed, as 
 shown by~\textcite{ku12superfluid} for Fermi gases at unitarity and by~\textcite{desb14}
 for Bose gases in two dimensions, the function $\tilde{\kappa}(p)$ is  
 accessible from precision measurements of in-situ density profiles in a harmonic trap. 
 In the following we will show that the stabilization of the gas right at zero scattering length 
 due to three-body interactions is sufficient to determine the universal behavior 
 near the quantum tricritical point which separates the liquid ground state in the regime $a<0$
 from the situation at $a>0$, where the ground state is a gas
and the phase diagram at finite temperature has the form shown in Fig.~\ref{fig:phase-diagram}. \\

\section{Zero temperature phase diagram and quantum tricritical point}

In order to derive an effective field theory of the zero temperature gas-liquid transition
in the many-body problem at a finite density $n$, 
we start from the microscopic action of a Bose system with pure two-body interactions 
as described by Eq.~(\ref{eq:Hamiltonian}).
The associated generating functional $Z[J]=\int D\psi\exp{(-S[\psi]/\hbar +\int\! J\psi)}$ for 
the correlation functions of the complex scalar field $\psi(\tau,\mathbf{x})$ can formally be expressed as a
functional integral with action   
\begin{equation}
 S=\int_{\tau}\, \int_{\mathbf{x}} \,\left\{ \psi^{\ast}(\tau,\mathbf{x}) \bigl(\hbar\partial_{\tau} -\frac{\hbar^2}{2m}\nabla^2-\mu\bigr)\psi(\tau,\mathbf{x})  +
 \frac{1}{2}|\psi(\tau,\mathbf{x})  |^2 \int_{\mathbf{x}'} \, V(\mathbf{x}-\mathbf{x}') |\psi(\tau,\mathbf{x}')  |^2 \right\}=S_0+S_{\rm int}\, .
 \label{eq:action}
 \end{equation}   
At the mean field level, the effective potential for a field configuration with no dependence on the time and spatial variables $\tau$ and $\mathbf{x}$,  
where $|\psi|^2=n$ can be identified with the particle density, has the form $V_{\rm eff}^{(0)}=-\mu n +(g/2)n^2$.
The coefficient $g\!=\! 4\pi\hbar^2\, a/m>0$ is fixed by the two-body scattering length in vacuum
\footnote{In the naive mean field approach, $g$ contains the scattering length in the Born approximation, which is ill-defined for
potentials which increase more strongly than $1/r^3$ at short distances. This problem is eliminated only in a full treatment of the 
two-body problem which is contained in the formulation in terms of an effective potential in Eq.~(\ref{eq:Coleman-Weinberg}) below.}, 
which is positive in the regime
$\Lambda_{\rm dB}>\Lambda_{\rm dB}^c$ where the two-body interactions are dominantly repulsive. The onset transition from the vacuum to a low 
density superfluid gas is then well understood in terms of a Gross-Pitaevskii description. In particular, the density of bosons 
$n(\mu)=\mu/g+\ldots$ rises linearly to lowest order as $\mu\to 0^+$, while $n(\mu)\equiv 0$ 
vanishes for negative values of the chemical potential. 
Thus, $\mu=0, g>0$ is a line of quantum critical points which separates the 
vacuum state from a superfluid gas at finite density~\cite{sach11book}. Despite the finite jump in the compressibility
from $\tilde{\kappa}=0$ to $\tilde{\kappa}=1/g>0$,  the vacuum to superfluid transition is a continuous
one. Indeed, approaching the line $\mu=0$ from positive values, the correlation length is the well known healing
length $\xi=\hbar/\sqrt{2m\mu}=(8\pi na)^{-1/2}$ of a weakly interacting Bose-Einstein condensate. It is large 
compared to the  average interparticle spacing since $na^3\ll 1$ in the low density limit.  
Moreover, using the zero temperature Gibbs-Duhem relation $\mu=u+p/n$ which connects the chemical potential 
and the pressure to the energy $u$ per particle, one has $u(n)=gn/2=\sqrt{gp/2}\to 0$ in the limit of vanishing density or 
pressure.\\

\begin{figure}[t]	
\includegraphics[width=70mm]{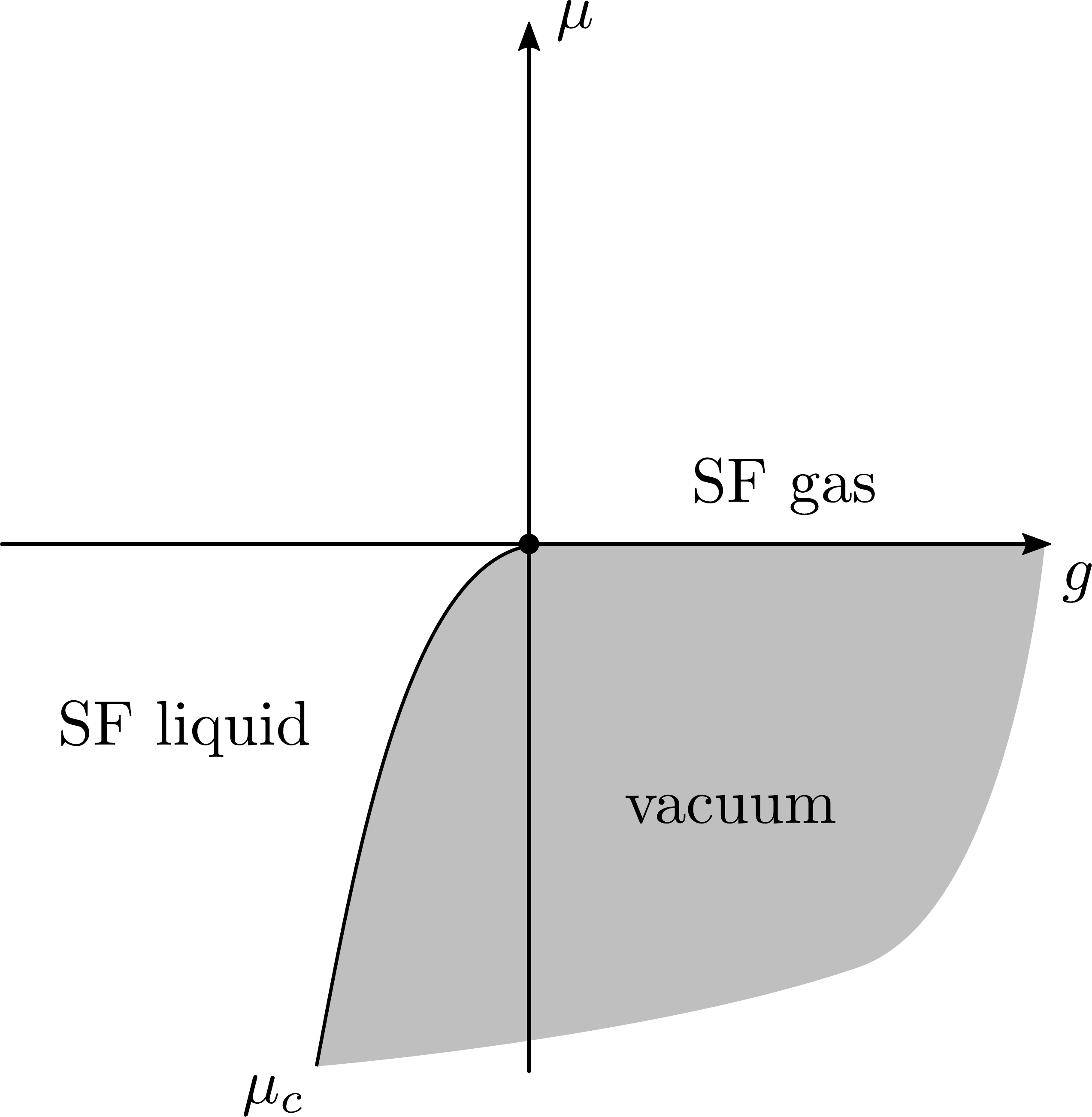}
\caption{Zero temperature phase diagram as a function of the chemical potential $\mu$ 
and the deviation $g\sim\Lambda_{\rm dB}-\Lambda_{\rm dB}^c$ of the de Boer parameter 
from its critical value. The gaseous ground state in the regime $g>0$ arises from the vacuum 
at $\mu<0$ via a continuous transition. For $g<0$, the ground state is a liquid. It is separated
from the vacuum by a first order transition at $\mu_c<0$. The point $\mu=g=0$ is a quantum tricritical point. }
\label{fig:tricritical}
\end{figure}

The endpoint at $g=0$ of the line $\mu\equiv 0$ is a quantum tricritical point (see Fig.~\ref{fig:tricritical}).
It separates the continuous onset transition from the vacuum to a gaseous state in the regime $g>0$ 
from a first order transition at $\mu_c<0$ between the vacuum and a liquid 
for negative values of the scattering length. Now, as shown by~\textcite{son06symmetry}, 
the  leading order effective field theory describing any finite density superfluid 
is completely fixed by the relation between the pressure and the chemical potential.
Specifically, using~(\ref{eq:critical-pressure}), the Lagrange density (expressed in real time)
of the superfluid just above the quantum tricritical point reads
\begin{equation}
 \mathcal{L}_{\rm eff}\vert_{g=0}=\left(\frac{8m}{9\hbar^2 D}\right)^{1/2}\!\left( \hbar\partial_t{\theta}-\frac{\hbar^2}{2m} (\nabla\theta)^2\right)^{3/2} \;\; \to \;\;
 p(\mu)\vert_{g=0} - n(\mu)\,\hbar\partial_t\varphi + \frac{\tilde{\kappa}(\mu)}{2}(\hbar\partial_t \varphi)^2-\frac{\hbar^2 n(\mu)}{2m}(\nabla\varphi)^2 +\ldots
 \label{eq:critical-EFT}
 \end{equation}
 Upon expansion to quadratic order in small gradients of the variable $\theta=\mu t/\hbar -\varphi(t,\mathbf{x})$, 
 it gives rise to the standard quantum hydrodynamic description of a superfluid in terms of 
 a time and space dependent phase variable $\varphi(t,\mathbf{x})$, see e.g.~\cite{pita16book}.
The velocity $c_s$ of the resulting phonon-like excitations is determined from
\begin{equation}
mc_s^2=n/\tilde{\kappa}\to 2\mu=\frac{\hbar^2 D}{m}\, n^2(\mu)\;\; {\rm at} \; g=0\, .
 \label{eq:sound-velocity}
 \end{equation}
In a superfluid gas at vanishing scattering length, therefore, the velocity $c_s$ depends linearly on the density rather than the standard square root
behavior found for positive scattering lengths $g>0$. \\

A more complex situation arises for $\Lambda_{\rm dB}< \Lambda_{\rm dB}^c$, where the scattering length
is negative.
In order to properly deal with the regime $g<0$, where the ground state at vanishing pressure is a finite density liquid, 
it is necessary to include the quantum fluctuations of the 
field $\psi(\tau,\mathbf{x})$ to all orders. On a formal level, this  can be expressed in terms of an effective potential
 \begin{equation}
 \Gamma[\psi]=\sum_{N=1}^{\infty} \frac{1}{N!} \int_{p_1\ldots q_N}\!\Gamma_N(p_1\ldots p_N\, q_1\ldots q_N)\, \psi^{\ast}(p_1)\ldots
 \psi^{\ast}(p_N)\, \psi(q_1)\ldots \psi(q_N) = \int_{\tau,\mathbf{x}}\!\left\{ V_{\rm eff}[\psi]+\psi^{\ast}\tilde{D}\psi +\ldots \right\}
 \label{eq:Gamma}
 \end{equation}  
 which is defined via a Legendre transform $ \Gamma[\psi]=\ln{\{Z[J]/Z[0]\}}-\int J\psi$ of the generating functional $Z[J]$ associated with
 the action~(\ref{eq:action})
 \footnote{Due to Galilei invariance, derivatives can only appear in the covariant form $\tilde{D}=\hbar\partial_{\tau}-\hbar^2\nabla^2/2m$.} .  
 It contains the exact vertex functions $\Gamma_N$ at arbitrary orders, which are essentially the amplitudes 
 for scattering processes with $N$ incoming and $N$ outgoing particles. Knowledge of the $\Gamma_N$,
 including their dependence on the $2N$ momentum variables $p_1\dots q_N$ which are constrained only by translation invariance in space and time
 $p_1+\ldots +p_N=q_1+\dots +q_N$, therefore requires a complete solution of the $N$-body problem. This is clearly impossible. 
 Fortunately, however, for the discussion of the behavior near the  quantum tricritical point, which is a zero density fixed point,
we need only the leading non-vanishing terms in the expansion of the effective potential 
\begin{equation}
V_{\rm eff}[\psi]= -\mu |\psi|^2+\frac{g}{2}\, |\psi|^4 + \frac{\lambda_3}{3} |\psi|^6  +\ldots
\label{eq:Coleman-Weinberg}
\end{equation}
associated with a time and space independent 'classical' field $\psi$.
Here, the coefficient $\lambda_3=\hbar^2 D/2m$ of the contribution $\sim |\psi|^6 $ arises from the zero momentum limit $\Gamma_3(0)=\hbar^2D/m$ 
of the vertex function which is associated with effective three-body interactions. It is fixed by the hypervolume $D$ discussed above 
and may in principle be calculated for a given two-body potential by solving for the three-body wave function at zero energy,
as specified in Eq.~(\ref{eq:hypervolume}).   
In the regime $g<0$, the symmetry broken phase with a finite density $n(\mu)=|\bar{\psi}|^2\ne 0$ appears
already beyond a negative value 
\begin{equation}
\mu_c=-3g^2/(16\lambda_3)=-6\pi^2\,\hbar^2a^2/(mD)
\label{eq:chemical-potential}
\end{equation}
of the chemical potential, which vanishes with the square of the distance 
from the quantum tricritical point as indicated in Fig.~\ref{fig:tricritical}. 
Remarkably, this behavior is identical to that found in the numerical 
approach to the bosonic many-body problem near  $\Lambda_{\rm dB}^c$ by~\textcite{mill77}.
Specifically, they have determined the dimensionless curvature 
\begin{equation}
\tilde{u}''_c : = 
\left. \frac{1}{4\epsilon_{\sigma}}\,\frac{d^2 u(p=0)}{d \Lambda_{\rm dB}^2}\right |_c
= - 3\pi^2 a_{\Lambda}^2\,\frac{\sigma^2\lvdW^2}{D}
\label{eq:curvature}
\end{equation}
of the energy per particle $u(p=0)$ right at the critical point. By the Gibbs-Duhem relation,
$u(p=0)=\mu_c$ coincides with the critical chemical potential since the pressure vanishes 
along the line separating the vacuum from the finite density liquid. 
In Eq.~(\ref{eq:curvature}), $u(p=0)$ has been normalized 
by the zero point energy $\epsilon_{\sigma}=\hbar^2/m\sigma^2$
at the scale $\sigma$, while $a_{\Lambda}$ is the numerical constant which
appears in Eq.~(\ref{eq:zero-crossing}).  Recalling that for a Lennard-Jones interaction 
one has $\lvdW=0.86\,\sigma$ at zero scattering length, the numerical result $\tilde{u}''_c\simeq -6.547$ 
of Ref.~\cite{mill77} for the dimensionless critical curvature leads to an associated three-body scattering
hypervolume $D(a=0)\simeq 3.34\,a_{\Lambda}^2\sigma^4$. The assumption of a positive value 
of the three-body scattering hypervolume near $a=0$ 
for interactions which are strongly repulsive below a short distance scale $\sigma$ is 
therefore supported implicitly by the numerical results of~\textcite{mill77}.  They did not
realize, however, the connection between $\Lambda_{\rm dB}^c$ and the zero of
the scattering length nor the relation between the finite curvature~(\ref{eq:curvature}) 
and the three-body problem through the associated scattering hypervolume.
The extraction above of its specific value near $a=0$ from numerical results of a many-body 
calculation thus obscures the fact that the parameter $D$ is fully determined by solving a three-body problem. 
Since $u(p=0)\equiv 0$ in the regime of positive scattering lengths, 
the second derivative of $u(p=0)$ with respect to $\Lambda_{\rm dB}$ 
exhibits a jump at the quantum tricritical point. 
In physical terms, this gives rise to a jump in the 
derivative of the kinetic energy per particle $u_{\rm kin}=(\Lambda_{\rm dB}/2)du/d\Lambda_{\rm dB}$,
which has been interpreted as a signature for a conventional second order transition~\cite{mill77}.
This conclusion, however, hides the presence of  
the underlying quantum tricritical point, whose critical exponents differ from those in standard Landau theory. Moreover,
the virial theorem along the zero pressure line  $\mu=\mu_c$, which takes the simple form $u_{\rm kin}+6u_{12}-3u_6=0$ 
for the specific example of a Lennard-Jones interaction~\cite{mill77}, becomes trivial at the quantum tricritical point 
because it is a zero density fixed point where both $u_{\rm kin}$ and $u_{\rm int}$ vanish. \\

Right on the line $\mu=\mu_c$, 
the density jumps from zero in the vacuum state $\mu<\mu_c$ to a finite value 
\begin{equation}
\bar{n}=n(\mu_c)=3|g|/(4\lambda_3)=6\pi\, |a|/D \;\;\; \to \;\;\; \bar{n}\sigma^3=6\pi\, |a|\sigma^3/D=
\frac{2|\tilde{u}''_c|}{0.86\,\pi a_{\Lambda}}\left(\Lambda_{\rm dB}^c - \Lambda_{\rm dB}\right) \, .
\label{eq:density}
\end{equation}
The dimensionless product $\bar{n}\sigma^3$ therefore approaches zero linearly with the 
deviation from the quantum tricritical point, with a numerical prefactor of order one. 
Despite the large deviation $\Lambda_{\rm dB}^c - \Lambda_{\rm dB}\simeq 0.28$ 
of the de Boer parameter from its critical value, one might naively try to use Eq.~(\ref{eq:density}) for 
a rough estimate of the density at zero pressure in $^4$He, whose empirical value is $\bar{n}\sigma^3=0.364$.
Such an estimate is misleading, however, because the $^4$He interaction supports a two-body bound state
and thus the relation~(\ref{eq:zero-crossing}) underlying this estimate does not apply. 
Quite generally, the equation of state in the liquid is of the form
\begin{equation}
p(\mu)=-V_{\rm eff}[n(\mu)]=\left[ \bar{n}\left(\mu-\mu_c\right)+\tilde{\kappa}_c\left(\mu-\mu_c\right)^2\!/2 +\ldots\right] \,\Theta\left(\mu-\mu_c\right)\, .
\label{eq:liquid-pressure}
\end{equation}
Since the compressibility $\tilde{\kappa}=\partial^2 p/\partial\mu^2$ is positive, $p(\mu)$ is a convex function of the chemical potential
which must vanish identically in the vacuum regime $\mu<\mu_c$. 
Here, a constant in~(\ref{eq:Coleman-Weinberg}) has been added to guarantee $V_{\rm eff}[\bar{n}]=0$. Moreover, there 
is a minus sign in comparison with the effective Lagrange density of Eq.~(\ref{eq:critical-EFT}) since the latter
is expressed in real time $t$ rather than the imaginary time variable $\tau$ used in Eq.~(\ref{eq:Gamma}). 
The density profile at a liquid-to-vacuum boundary with an effective potential of the form~(\ref{eq:Coleman-Weinberg}) 
has been calculated by~\textcite{bulg02}. It has the form $n(z)=\bar{n}/\left( 1+\exp(2\kappa_0 z)\right)$ with a 
healing length $1/\kappa_0=\hbar/\sqrt{2m|\mu_c|}\simeq \sqrt{D}/|a|$ which diverges linearly with the distance from the quantum tricritical point. 
As a result, the surface tension derived in Ref.~\cite{bulg02},  
\footnote{We use a bar in the surface tension $\bar{\sigma}$ to distinguish it from the short distance length scale $\sigma$. Note also
that the exponent $\nu_u=\nu_t/\phi_t=1$ for the divergence of the correlation length $1/\kappa_0$ along the first-order transition line $\mu=\mu_c$
is a subsidiary tricritical exponent in the notation of~\textcite{grif73}. The relevant crossover exponent $\phi_t=1/2$ is determined by the quadratic 
behavior~(\ref{eq:chemical-potential}) of the chemical potential near the quantum tricritical point.} 
\begin{equation}
\bar{\sigma}=\frac{\lambda_3\bar{n}^3}{6\,\kappa_0}\simeq \frac{\hbar^2\, a^2 }{m\, D^{3/2}}
\label{eq:tension}
\end{equation}
vanishes quadratically $\bar{\sigma}\sim \left(\Lambda_{\rm dB}^c - \Lambda_{\rm dB}\right)^2$,
consistent with a scaling relation due to~\textcite{wido65} which connects the exponent of the 
surface tension $\bar{\sigma}\sim 1/\xi^{d-1}$ with that of the correlation length. 
This result will play a crucial role in the following section, where we discuss the shift of the zero temperature 
liquid-gas transition for finite particle numbers. \\

Finally, we mention that within the microscopic model defined by Eq.~(\ref{eq:action}),
there are in fact {\bf two} separate quantum phase transitions which occur as a function 
of the de Boer parameter $\Lambda_{\rm dB}$. The first one, discussed here, appears 
between a gaseous and a liquid ground state 
at a critical value $\Lambda_{\rm dB}^c\simeq 0.7$ where the scattering length crosses zero. 
Decreasing the de Boer parameter further below values $\Lambda_{\rm dB}\simeq 0.4$
 characteristic for $^4$He, the liquid ground state will eventually turn into a solid 
via a first order quantum phase transition. On the basis of a variational Ansatz for the 
ground state wave function, the associated critical de Boer parameter for bosons 
has been estimated to be around $\Lambda_{\rm dB}^{\rm c, solid}\simeq 0.37$ 
by~\textcite{nosa75}.  In the case of Fermions  
$\Lambda_{\rm dB}^{\rm c, solid}\vert_F\simeq 0.42$ is substantially larger 
because Fermions prefer to stay localized near a discrete set of lattice 
sites even for larger values of the zero point motion. Note that,  
in contrast to the gas-liquid transition studied here, these critical values 
cannot be determined from two-body physics. \\

\section{Quantum unbinding for finite particle numbers}

In the regime $\Lambda_{\rm dB}<\Lambda_{\rm dB}^c$ of negative scattering 
lengths, the ground state at vanishing pressure is a superfluid liquid. By the
Gibbs-Duhem relation, the energy per particle $u(p=0)=\mu_c<0$ is negative.
A given number $N$ of particles thus has an extensive binding energy
$B_N=|u(p=0)|\, N$. Moreover, since the liquid has a finite
density $\bar{n}$ at zero pressure, the radius of the associated bound state scales like 
$R_N\simeq (N/\bar{n})^{1/3}$. In the limit where the scattering length 
approaches zero, both $u(p=0)$ and $\bar{n}$ vanish.
The zero pressure liquid thus evaporates into a gas precisely at the quantum 
tricritical point $\mu=g=0$. 
This is true, however, only in the thermodynamic limit. For finite particle numbers,
the binding energy $B_N$ is reduced because particles on the surface of the
associated cluster are less bound than those in the bulk. 
For the specific case of a Lennard-Jones interaction, this has been studied 
numerically for small clusters by~\textcite{meie96} 
and by~\textcite{sevr10}. In particular, it has been found that, at finite $N$,
quantum unbinding appears at values $\Lambda_{\rm dB}^{\ast}(N)<\Lambda_{\rm dB}^c=0.679...$ 
of the de Boer parameter which are considerably lower than what is expected
in the thermodynamic limit. This observation can be understood 
by including a finite, positive surface energy $f_s$ per particle in the liquid phase,
which also accounts for the essentially flat radial density distributions
found numerically near $\Lambda_{\rm dB}^c$~\cite{sevr10}.  
The surface energy is defined by the subleading term in the expansion 
\begin{equation}
E_0(N)=u\, N + f_s\, N^{2/3}+\ldots
\label{eq:surface}
\end{equation}
of the $N$-body ground state energy for $N\gg 1$.
In practice, this expansion requires particle numbers larger than $N\simeq 10$. 
In fact, numerical studies of small Helium clusters, where the scattering
length is close to infinity rather than $a\simeq 0$ as studied here, yield $E_0(N)\sim N^2$
up to $N=10$, see e.g.~\textcite{yan15N-body}.
Based on the finite size generalization of the liquid ground state energy in Eq.~(\ref{eq:surface}), 
the unbinding condition $E_0(N+1)=E_0(N)$ of a vanishing single particle 
addition energy $\mu(N)=E_0(N+1)-E_0(N)=0$ can be written in the form
\begin{equation}
\frac{-3\, u}{\,2  f_s}\left[\Lambda_{\rm dB}^{\ast}(N)\right]=N^{-1/3} \, .
\label{eq:unbinding}
\end{equation}
The finite size scaling of the deviation $\Lambda_{\rm dB}^c-\Lambda_{\rm dB}^{\ast}(N)$
for $N\gg 1$ is thus determined by the dependence of the bulk energy $u$ and the surface 
energy $f_s$ per particle on the de Boer parameter.  In Ref.~\cite{sevr10} it has been assumed
that $u(\Lambda_{\rm dB})$ \newline vanishes linearly near $\Lambda_{\rm dB}^c=\Lambda_{\rm dB}^{\ast}(\infty)$. 
This leads to $\Lambda_{\rm dB}^{\ast}(\infty)-\Lambda_{\rm dB}^{\ast}(N)\sim N^{-1/3}$
provided $f_s$ is finite at the quantum tricritical point. This is clearly not the case, however,
because the ground state at $g=0$ is a gas, which has vanishing surface energy.
In order to determine the proper scaling for large $N$, we use the result~(\ref{eq:chemical-potential}) 
derived above. It shows that
the energy per particle $u(p=0)=\mu_c$ on the zero pressure line separating the vacuum from the finite density liquid
vanishes like the square of the deviation from the quantum tricritical point. 
Indeed, the quadratic dependence near $\Lambda_{\rm dB}^c$ agrees quite well 
with the numerically calculated ground state energy of~\textcite{mill77}
(see their Fig. 4). \newline It also provides a much better extrapolation towards the critical point of the data shown in 
Figures 2 b) and 10 of Ref.~\cite{sevr10} than the assumption of a linear behavior. 
To determine how the surface energy $f_s$ per particle vanishes near
$\Lambda_{\rm dB}^c$,  we follow an argument due to~\textcite{bulg02}, who has used the 
expansion~(\ref{eq:surface}) to discuss Bose droplets with $N$ particles in the vicinity of the 
scattering lengths where the three-body scattering hypervolume $D$ diverges.
For the conceptually quite different situation of small negative scattering lengths and no three-body bound 
states discussed here, the results~(\ref{eq:density}) for the average interparticle spacing 
$\bar{n}^{-1/3}$ and~(\ref{eq:tension}) for the surface tension imply that
the surface energy $f_s\simeq 4\pi\,\bar{n}^{-2/3}\cdot\bar{\sigma}\sim |\Lambda_{\rm dB}-\Lambda_{\rm dB}^c|^{4/3}$ 
vanishes with a nontrivial power law near the  quantum tricritical point. Based on Eqs.~(\ref{eq:unbinding}) and~(\ref{eq:chemical-potential}), 
the threshold values $\Lambda_{\rm dB}^{\ast}(N)$ of the de Boer parameter beyond which $N$-body 
bound states disappear therefore approach the critical value $\Lambda_{\rm dB}^c$ of the bulk liquid-gas transition according to
\footnote{In Ref.~\cite{sevr10}, a power law with an exponent $1/3$ has been fitted to the data for particle numbers between $N=2$ 
and $N=40$. In view of the restricted range in $N^{-1/3}$ the agreement found there is still consistent with the different 
asymptotic behavior predicted here.  The result~(\ref{eq:scaling}) also differs from an earlier analysis of 
the quantum unbinding problem by~\textcite{hann06},
where numerical results on Helium clusters were extrapolated assuming a much faster approach 
$\Lambda_{\rm dB}^c-\Lambda_{\rm dB}^{\ast}(N)\sim 1/N$ to the critical de Boer parameter. \\
}  
\begin{equation}
\Lambda_{\rm dB}^c-\Lambda_{\rm dB}^{\ast}(N)  \sim N^{-1/2}\, .
\label{eq:scaling}
\end{equation}
Moreover, in view of Eq.~(\ref{eq:zero-crossing}), this leads immediately to a power law behavior 
\begin{equation}
-a_{-}(N\gg 1)\simeq \left(\sqrt{D}/ N\right)^{1/2}
\label{eq:length-scaling}
\end{equation} 
of the associated scattering lengths.
Together with the basic conceptual structure of the phase diagram shown in Fig.~\ref{fig:tricritical},
the result~(\ref{eq:length-scaling}) is a major prediction of the present work. 
It has the remarkable feature that the three-body
scattering hypervolume $D(a=0)$ at vanishing scattering length sets the scale for 
the unbinding of $N$-body bound states in the asymptotic limit $N\gg 1$.
This is a consequence of the fact that $D$ appears in the leading term $\sim D\,|\psi|^6$ in 
Eq.~(\ref{eq:Coleman-Weinberg}) which stabilizes the superfluid at both vanishing
and small negative scattering lengths,  
while higher order contributions are negligible near the quantum tricritical point, where $\bar{n}\to 0$. 

\section{Conclusion}

To conclude, we have discussed the quantum phase transition between a 
liquid and a gaseous phase of bosons with increasing strength of the
zero point motion. A major result is encoded in Fig.~\ref{fig:tricritical}
which shows that, at vanishing pressure, these two phases are
separated by a quantum tricritical point. Its location is determined
by the condition $a(\Lambda_{\rm dB}^c)=0$ of a vanishing scattering length. 
Note that a similar type of phase diagram
was found previously by~\textcite{niko07renorm} for Fermi gases near unitarity.
In this case,  the boundary between the vacuum and a superfluid phase at finite 
density exhibits a quantum {\bf multi}critical point where two lines of fixed points
with continuous transitions meet. The present approach for dealing 
with the zero temperature liquid-gas transition for bosons is complementary 
to the early work on this problem by~\textcite{mill77}, which was based on
microscopic model calculations. As a result, the underlying universality and in
particular the  connections of the many-body problem with two- and three-body
physics was not realized. Indeed, apart from the fact that the location of the
quantum critical point is determined by two-body physics, it is a unique feature
 of the present problem that even the universal behavior in its vicinity is fixed by the 
 parameter $D$ which is fully determined by solving a three-body problem. In practice, the only 
 two systems where the physics discussed here is applicable in direct form are $^4$He and 
 spin-polarized hydrogen. In the latter case, the ground state is a superfluid gas with a finite 
 temperature phase diagram as shown in Fig.~\ref{fig:phase-diagram}. In turn, the  $^4$He 
 ground state is a superfluid liquid which, as noted above, is unfortunately quite far from a 
 system with small negative scattering length. 
Our results are of direct relevance, however, for numerical studies of Bose clusters 
near $\Lambda_{\rm dB}^c$, as performed by~\textcite{meie96} and by~\textcite{sevr10}.
In particular, a sytematic study of larger particle numbers should reveal the finite size scaling~(\ref{eq:scaling})
and also the associated dependence  $-u(\Lambda_{\rm dB}^{\ast}(N))\sim 1/N$ of the negative shift in 
energy per particle where clusters unbind at finite $N$,
which is based on our analysis of the ratio $u/f_s$ as a function of $\Lambda_{\rm dB}$. 
From a mathematical point of view, an open problem is to specify precisely the class of two-body 
interactions $V(\mathbf{x})$ for which the three-body scattering hypervolume $D$ is positive near the 
zero of the scattering length where no two-body bound states are present.\\
 
 A promising but more complex case for an application of the present ideas appears in ultracold gases, 
 where the scattering length and thus the effective de Boer parameter can be changed externally via 
 Feshbach resonances~\cite{chin10feshbach}.  
 In particular, it is straightforward to tune an ultracold Bose gas to zero scattering length.
  In contrast to the situation discussed in the context of Fig.~\ref{fig:deBoer}, however, the three-body scattering hypervolume 
now has a finite imaginary part associated with three-body losses via the relation~(\ref{eq:Bose-lifetime}).
 Moreover, since the de Boer parameter $\Lambda_{\rm dB}$ associated with the full two-body 
 interaction is much less than one, one has $\lvdW\gg\sigma$ and a true ground state which is a solid for both 
 signs of the scattering length. Yet, in the regime of non-negative scattering lengths $a\geq 0$, dilute ultracold 
 Bose gases stay in an effectively thermalized gaseous phase on time scales shorter than the inverse three-body
 loss rate $\Gamma_3^{-1}$. In particular, provided that the real part of the three-body scattering hypervolume $D$ 
is positive near the relevant zero crossing of the scattering length,
the equation of state is dominated by three-body interactions. The gas at finite density is thus described by an effective 
potential of the form~(\ref{eq:Coleman-Weinberg}) with $g=0$. As a result, it will exhibit a 
nontrivial relation~(\ref{eq:critical-kappa}) between compressibility and pressure and a
sound velocity which depends linearly on density, as derived in Eq.~(\ref{eq:sound-velocity}).
These relations provide a means to determine the parameter $D$ and it would
obviously be of interest to check these predictions experimentally.  A first step in this direction 
has been taken by~\textcite{shot14}, who have measured the recombination length $L_m$ 
in ${\rm Im}\, D\simeq L_m^4$ near a zero crossing of the scattering length at 
$B\simeq 850\,$G in $^7$Li.  Consistent with the dimensional argument above in the 
context of the purely real value $D\simeq (\ell_{\rm vdW})^4$ of the three-body scattering hypervolume 
near the zero crossing of the scattering length where no two-body bound state exists,  
the observed recombination length near $a=0$ turns out to be of order $L_m\simeq 4\,\ell_{\rm vdW}$~\cite{shot14}. 
For negative scattering lengths, the true ground state of single component ultracold Bose gases is a solid. 
The finite density liquid discussed in connection with Eq.~(\ref{eq:density}) in the regime 
near the critical de Boer parameter  $\Lambda_{\rm dB}^c$ is therefore not accessible. 
However, as shown by~\textcite{petr14}, a possible realization of a dilute liquid phase of bosons at negative scattering length 
which is stabilized by repulsive three-body interactions may be achieved in a situation where two internal 
states $|\!\uparrow\rangle$ and $|\!\downarrow\rangle$ are coupled by an rf-field. By varying the effective Rabi coupling,
the scattering length in the symmetric configuration $(|\!\uparrow\rangle + |\!\downarrow\rangle)/\sqrt{2}$ can be tuned
to zero. The associated three-body scattering hypervolume $D(a=0)\simeq a_{\uparrow\uparrow}^4/\xi$ 
is large and positive provided $\xi=(a_{\uparrow\downarrow}+a_{\uparrow\uparrow})/(a_{\uparrow\downarrow}-a_{\uparrow\uparrow})\ll 1$. 
In particular, it is a factor $1/\xi\gg 1$ larger than the characteristic magnitude 
${\rm Im}\, D\simeq a_{\uparrow\uparrow}^4$ of its imaginary part, which determines the standard scaling of the three-body loss rate. 
Neglecting losses, the resulting effective potential~(\ref{eq:Coleman-Weinberg}) gives rise to a dilute
Bose liquid in the regime where $a<0$. Its dimensionless density $\bar{n}a_{\uparrow\uparrow}^3\simeq \xi\,|a|/a_{\uparrow\uparrow}$ vanishes linearly
with the effective scattering length as in Eq.~(\ref{eq:density}) and - moreover - is small enough to be accessible in the
extremely dilute regime of ultracold gases. In fact, this type of liquid is a three-body interaction analog of self-bound droplets in two 
component Bose gases which are stabilized by the Lee-Huang-Yang contribution to the interaction energy, 
as predicted by~\textcite{petr15} and observed experimentally by~\textcite{cabr18}. A study of the unbinding 
of such droplets in the limit $a\to 0^-$ might open the possibility to verify the predictions from the finite size scaling analysis of the
disappearance of $N$-body bound states in section IV, complementary to the quite challenging    
extension of experimental data on loss features beyond $N=5$~\cite{zene13}. \\

{\bf Acknowledgements:} It is a pleasure to thank Sergej Moroz, Richard Schmidt and Robert Seiringer 
for constructive comments. I am particularly grateful to Dmitry Petrov for pointing out the paper by~\textcite{bulg02}
and an inconsistency in the definition of the correlation length and the associated surface energy in the 
original version of the manuscript. Moreover, I would like to acknowledge Philippe Nozi\`eres, whose 
Lecture Notes on 'Liquides et Solides Quantique' from a course at the Coll\`{e}ge de France in 1983
have provided an important part of the motivation for this project. The work has been completed during a stay at Nordita
in a program on 'Effective Theories of Quantum Phases of Matter' whose support is gratefully 
acknowledged.

\vspace*{-0.3cm}
\bibliography{References}
\end{document}